\documentclass[%
 reprint,
 amsmath,amssymb,
 aps,
 pra,
]{revtex4-2}
\usepackage{amsthm}
\usepackage{amsmath}
\usepackage{amsfonts}
\usepackage{amssymb}
\usepackage{graphicx}
\usepackage[margin=0.5in]{geometry}
\usepackage{float}
\usepackage{caption}
\usepackage{subcaption}
\usepackage{tikz-cd} 
\usepackage{xcolor}
\usepackage{ulem}

\begin{document}
\preprint{APS/123-QED}

\title{Asymptotic hydrographs and anomalous dispersion in mass-conserving storage cascades}

\author{Henrique Santos Lima$^1$, M\'{a}rk Honti$^2$, and Bal\'{a}zs S\'{a}ndor$^2$}
\affiliation{$^1$ Centro Brasileiro de Pesquisas Físicas, Rua Xavier Sigaud 150, Rio de Janeiro 22290-180, Brazil \\ $^2$ Budapest University of Technology and Economics, Department of Hydraulic and Water Resources Engineering, M\"{u}egyetem rakpart 3, Budapest 1111, Hungary}

\begin{abstract}
Sums of independent exponential random variables lead to the Erlang distribution, providing a direct probabilistic route from exponential waiting times to the integer-shape gamma law. This paper investigates how this classical construction changes when the exponential waiting-time density is replaced by the $q$-exponential density of nonextensive statistics. Our main result is an analytical asymptotic expression for the outflow of a mass-conserving cascade of reservoirs driven by a $q$-exponential waiting-time kernel. In the critical case $q=5/3$, the large-cascade flow rate converges to a stable L\'{e}vy density whose time argument is shifted by a Galilean-type transformation. This shifted L\'{e}vy law gives the asymptotic hydrograph of the cascade. We also found that for the entire regime $1<q<2$ the macroscopic dynamics  are governed by $\alpha$-stable L\'{e}vy laws. This proves that anomalous non-Gaussian dispersion can emerge from pure mass-conserving convolutional chains without invoking fractional derivatives.
\end{abstract}

\maketitle

\section*{Introduction}

The Poisson process is a central construction in applied probability, nonequilibrium physics, and transport theory. For a process with a constant event rate, exponential waiting times, Erlang waiting times for a fixed number of events, and Poisson event counts represent three equivalent mathematical descriptions of the same renewal mechanism. This equivalence is also the mathematical basis of the classical linear cascade: the response of a sequence of identical reservoirs is obtained by repeatedly convolving the same exponential waiting-time density, producing an Erlang hydrograph. The situation changes fundamentally when the elementary waiting time is no longer exponential. Then the convolutional structure remains required by mass conservation, but the resulting cascade response need not belong to the same parametric family as the one-step kernel.

This observation is particularly important in non-ideal transport. Watersheds, industrial reactors, porous media, and heterogeneous environmental systems often contain stagnant regions, bottlenecks, multiscale storage zones, and broad spectra of relaxation rates. These mechanisms generate long waiting times and algebraic tails, so that the exponential kernel used in the classical Nash cascade becomes too restrictive~\cite{Nash1957,Nash1959,Dooge1959,RodriguezIturbe1979,SzollsiNagy1982,Szilagyi2006,Singh1997,Singh1976,SinghWoolhiser2002}. A central question is therefore not only which heavy-tailed kernel should replace the exponential one, but also what analytical form emerges after many mass-conserving convolutional stages.

The $q$-exponential formalism provides a natural framework for this problem. It originates in nonextensive statistical mechanics~\cite{Tsallis1988,Tsallis2023,Borges2004} and has been connected to anomalous diffusion, L\'evy statistics, and generalized central-limit behavior~\cite{Tsallis1995,Umarov2008,Lutz2003}. In this setting, the $q$-exponential waiting-time density is expressed as:
\begin{equation}
\psi(t)=(2-q)\lambda[1-(1-q)\lambda t]_{+}^{\frac{1}{1-q}}\,,\quad t\ge0\,,
\end{equation}
where $\lambda>0$, $1<q<2$, and $[x]_{+}\equiv\max\{x,0\}$ denotes the positive part of $x$.  This expression reduces to the ordinary exponential density as $q\to1^+$ and develops an algebraic power-law tail for $q>1$. Recent investigations have revealed that such power-law behaviors can emerge naturally from local, discrete-time Markovian dynamics where a spatial feedback mechanism induces strong effective correlations along the trajectories~\cite{LimaCurado2026}. Related algebraic scaling also appears in critical thermodynamic response parameters~\cite{Soares2025}, anomalous heat transport and breakdowns of Fourier's law~\cite{LimaTsallis2023,LimaTsallis2026}, and superstatistical descriptions of turbulent circulation fluctuations~\cite{LimaPereira2026}.

In this paper, we derive the large-cascade asymptotics of a mass-conserving reservoir chain whose waiting-time kernel is $q$-exponential. This provides an analytical expression for the flow rate after many convolutional stages, rather than only a numerical or phenomenological approximation. An important result arises in the critical case $q=5/3$. This value is physically motivated by spectral mechanisms in layered diffusion systems, where asymmetric volume initialization projects onto low-wavenumber eigenmodes and generates an emergent Gamma distribution of decay rates, yielding the $q$-exponential decay function exactly~\cite{Nazari2021,SandorHontiLima2026,Tsallis1995,Plastino1995,Borland1998,Bologna2000}. In this case, the small-$s$ expansion of the Laplace transform of the cascade outflow leads to a stable L\'{e}vy density. The subleading linear term in $s$ produces a Galilean-type shift of the time variable, so that the asymptotic flow rate is not merely a standard L\'{e}vy law but a shifted L\'{e}vy hydrograph. This result identifies the exact long-cascade transport kernel generated by repeated $q$-exponential waiting times. We also show that the transport dynamics  across the entire regime $1<q<2$ are in robust agreement with shifted asymmetric $\alpha$-stable L\'{e}vy laws.

\section*{Model and methods}

We formulate a uniform cascade used to connect the convolutional distribution with a mass-conserving transport model. Consider the system
\begin{equation}
\frac{dS_j}{dt}=I_j(t)-O_j(t)\,,\quad j=1\dots k
\label{c2}
\end{equation}
which describes the temporal evolution of stored volume $S_j$ in the $j$-th element of the cascade in terms of the outflow $O_j(t)$ and inflow $I_j(t)$ volume flow rates. The waiting-time density $\psi(t)$ connects the outflow to the inflow as $O_j(t)=\int_0^t I_j(\tau)\psi(t-\tau)\,\mathrm{d}\tau$. For a cascade of $k$ reservoirs with a Dirac-delta impulse at the first element, the system of mass conservation equations reads as
\begin{align}
\frac{dS_1}{dt}&=-O_1\,,\nonumber\\
\frac{dS_j}{dt}&=O_{j-1}-O_{j}\,,\quad j=2\dots k\,.
\label{e2}
\end{align}

Generating the outflows yields $O_1(t)\propto\psi(t)$ and $O_j(t)\propto*^j\psi(t)$ for $j=2\dots k$. Consequently, Eqs.~(\ref{e2}) can be written as
\begin{align}
\frac{dS_1}{dt}&\propto-\psi\,,\nonumber\\
\frac{dS_j}{dt}&\propto*^{j-1}\psi-*^j\psi\,,\quad j=2\dots k\,,
\label{e4}
\end{align}
where $*^1\psi\equiv\psi$.

With $S_1(0)=1$ and $S_j(0)=0$ for $j=2\dots k$, the Laplace transform of Eqs.~(\ref{e4}) is
\begin{align}
\tilde{S}_1(s)&\propto\frac{1-\tilde{\psi}(s)}{s}\,,\nonumber\\
\tilde{S}_j(s)&\propto\frac{[\tilde{\psi}(s)]^{j-1}\big(1-\tilde{\psi}(s)\big)}{s}\,,\quad j=2\dots k\,.
\end{align}
This system of equations provides a consistent framework for flow simulations via a chain of reservoirs for uniform waiting time densities, satisfied mass conservation.

\section*{Results}

\subsection*{Explicit asymptotic derivation for all possible values of $q$}

Let us derive the explicit asymptotic expressions for the outflow $O_k\propto\tilde{\psi}^k(s)$ for all admissible regimes, this is to say, $1<q<2$. The Laplace transform of the $q$-exponential waiting-time density is given by
\begin{equation}
\tilde{\psi}(s)= \frac{2-q}{q-1} \left[\frac{s}{(q-1)\lambda}\right]^{\frac{2-q}{q-1}} e^{\frac{s}{(q-1)\lambda}} \Gamma\left(\frac{q-2}{q-1},\,\frac{s}{(q-1)\lambda}\right),
\end{equation}
where $\Gamma(\cdot,\cdot)$ denotes the incomplete gamma function \cite{Olver2010}. Introducing the dimensionless variables $z = s / ((q-1)\lambda)$ and $\alpha = (2-q)/(q-1)$, we exploit the series expansion of the incomplete gamma function as
\begin{equation}
    \Gamma(-\alpha, z) = -\frac{\Gamma(1-\alpha)}{\alpha} + z^{-\alpha}\sum_{n=0}^\infty\frac{(-1)^n z^n}{n!(\alpha-n)}\,.
\end{equation}
Consequently, neglecting terms of order higher than one, the expansion yields
\begin{equation}
    \tilde{\psi}(z) \approx 1 - \Gamma(1-\alpha)z^\alpha - \frac{z}{\alpha-1}
\end{equation}
for $\alpha \neq 1$. For the boundary case $\alpha = 1$, the relation $\Gamma(-1,z) = z^{-1}e^{-z} - \Gamma(0,z)$ holds. Utilizing the connection between $\Gamma(0,z)$ and the exponential integral $E_1(z)$, we write
\begin{equation}
   E_1(z) = -\gamma - \ln{z} - \sum_{n=1}^\infty \frac{(-1)^n z^n}{n! n}\,,
\end{equation}
where $\gamma$ is the Euler-Mascheroni constant. Neglecting higher-order terms leads to the low-frequency approximation
\begin{equation}
    \tilde{\psi}(z) \approx 1 + z\ln{z} + \gamma z\,.
\end{equation}

For $\alpha \neq 1$, the logarithmic generator of the cascade response, $\ln{\tilde{\psi}^k(s)}$, expands as
\begin{equation}
    \ln{\tilde{\psi}^k(s)} \approx k \ln{\left(1 - z^\alpha \Gamma(1-\alpha) - \frac{z}{\alpha-1}\right)}\,.
\end{equation}

Assuming $1/3 < \alpha \leq 1/2$, the second-order term of the logarithmic expansion introduces lower-order structural modifications, such that
\begin{equation}
    \ln{\tilde{\psi}^k(s)} \approx k\left[-\Gamma(1-\alpha)z^\alpha - \frac{z}{\alpha-1} - \frac{(\Gamma(1-\alpha))^2 z^{2\alpha}}{2}\right] .
\end{equation}
Note that if $\alpha < 1/p$, where $p$ is a positive integer, additional lower-order terms emerge in the power series. For instance, if $p = 3$ or $p = 4$, third-order contributions from the logarithmic expansion become non-negligible, and a fourth-order contribution appears when $p = 4$. Fortunately, the asymptotic tail behavior is entirely determined by the lowest-order contribution; higher-order terms are required only to refine the solution at shorter timescales.

Conversely, when $\alpha > 1/2$, the expansion simplifies to
\begin{equation}
    \ln{\tilde{\psi}^k(s)} \approx k\left[-\Gamma(1-\alpha)z^\alpha - \frac{z}{\alpha-1}\right] .
\end{equation}
Alternatively, setting $\alpha = 1$ yields
\begin{equation}
    \ln{\tilde{\psi}^k(s)} \approx k\left[z\ln{z} + \gamma z\right] .
\end{equation}
Given that $z = (\alpha+1)s/\lambda$ when $\alpha=1$, substituting this into the expansion yields
\begin{equation}
    \ln \tilde{\psi}^k(s) \approx \frac{2k}{\lambda} s \ln s + \left[ \frac{2k}{\lambda} \left( \gamma + \ln\left(\frac{2}{\lambda}\right) \right) \right] s\,.
\end{equation}
Although the coefficients in this expression appear positive, which would superficially suggest exponential growth, evaluating the low-frequency regime $s \ll 1$ reveals that $s\ln{s}$ is strictly negative. This structure corresponds to the characteristic function of the Landau distribution \cite{Landau1944}. 

\begin{figure*}
    \centering
    \includegraphics[width=0.7\linewidth]{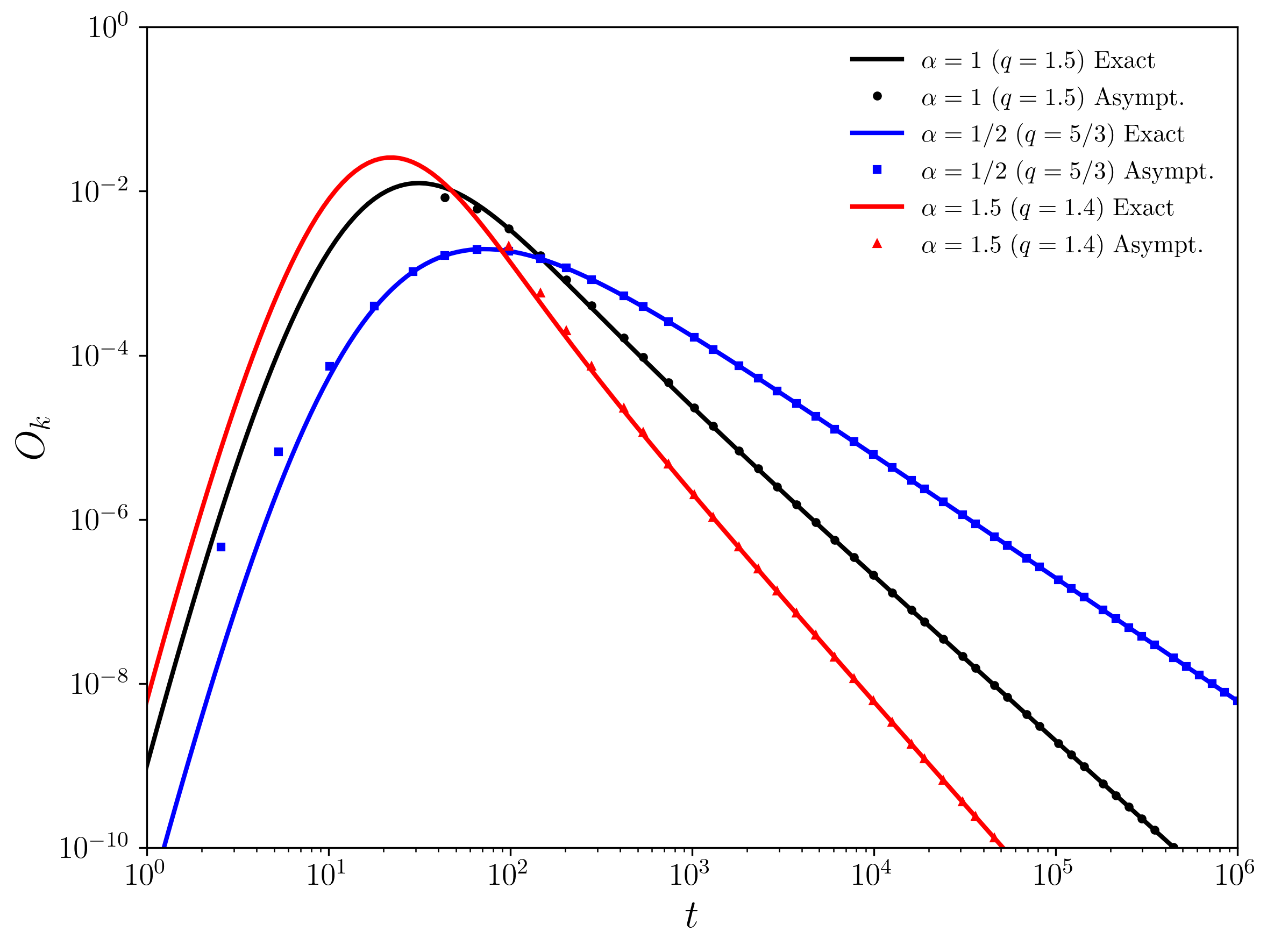}
    \caption{Log-log comparison of the exact $k$-fold convolutional outflow with the corresponding asymptotic prediction for $k=10$. Solid curves denote the numerical convolutions, and points denote the asymptotic formulas for three representative regimes: $\alpha=1$ ($q=3/2$), $\alpha=1/2$ ($q=5/3$), and $\alpha=1.5$ ($q=1.4$). For $\alpha=1$ and $\alpha=1.5$, points are shown only for $t\ge 5\times10^1$ because of numerical limitations at shorter times.}
    \label{fig:alpha-regime-comparison}
\end{figure*}

Applying the inverse Laplace transform, the large-cascade outflow decays as $O_k(t) \sim A t^{-(\alpha+1)}$ across all parametric regimes (where $A>0$), matching the asymptotic tail behavior of asymmetric $\alpha$-stable Lévy distributions \cite{Chambers1976, Nolan1997, Gorska2011}. From this power-law scaling, we conclude that the cumulative storage density follows $S_k(t) \sim A t^{-\alpha}$. FIG.~\ref{fig:alpha-regime-comparison} illustrates this result for representative values of $\alpha$, showing that the asymptotic laws reproduce the long-time slopes of the exact convolutional outflows.

\subsection*{The critical case $q=5/3$}

\begin{figure*}
\centering
\includegraphics[width=0.55\linewidth]{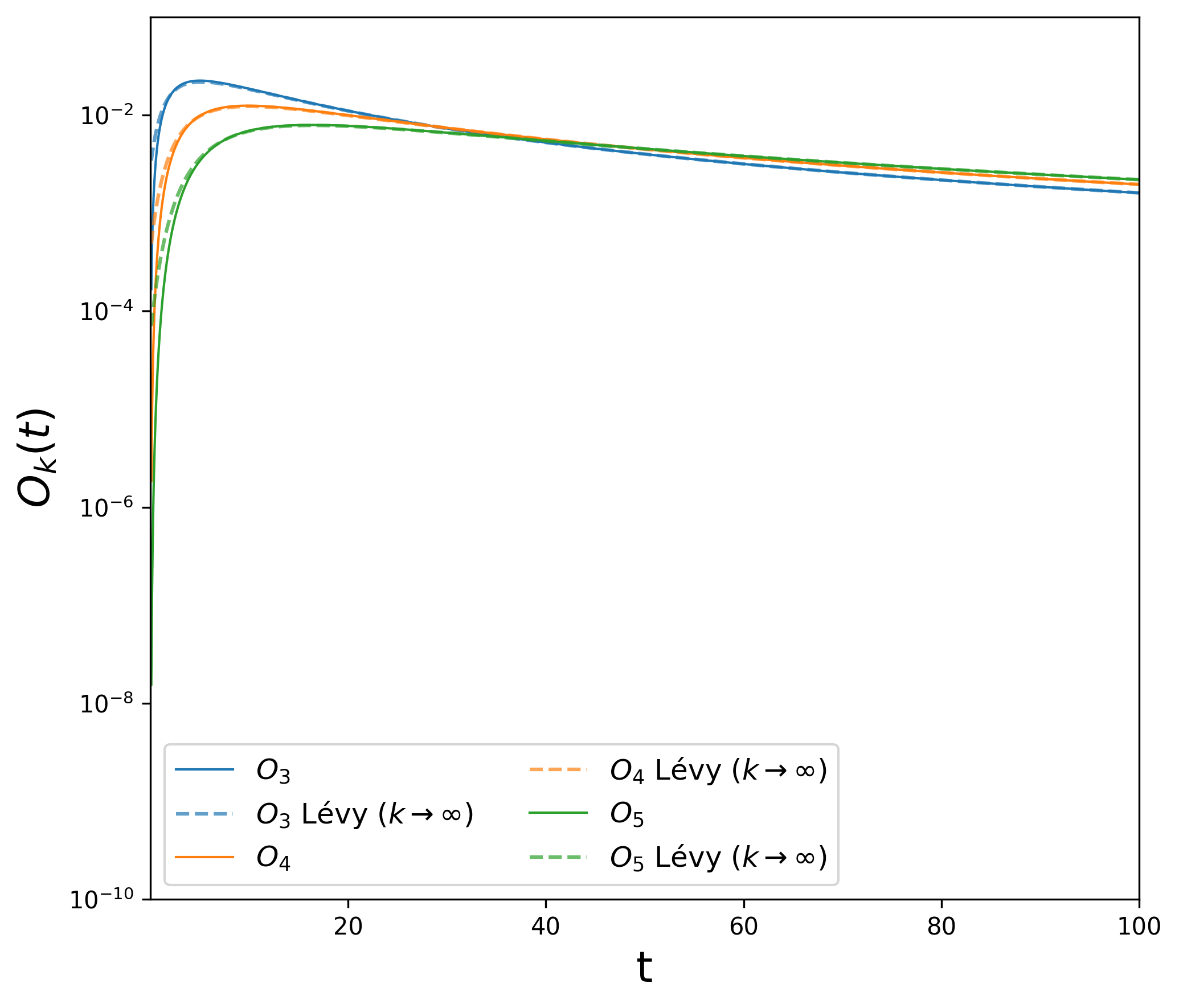}
\includegraphics[width=0.55\linewidth]{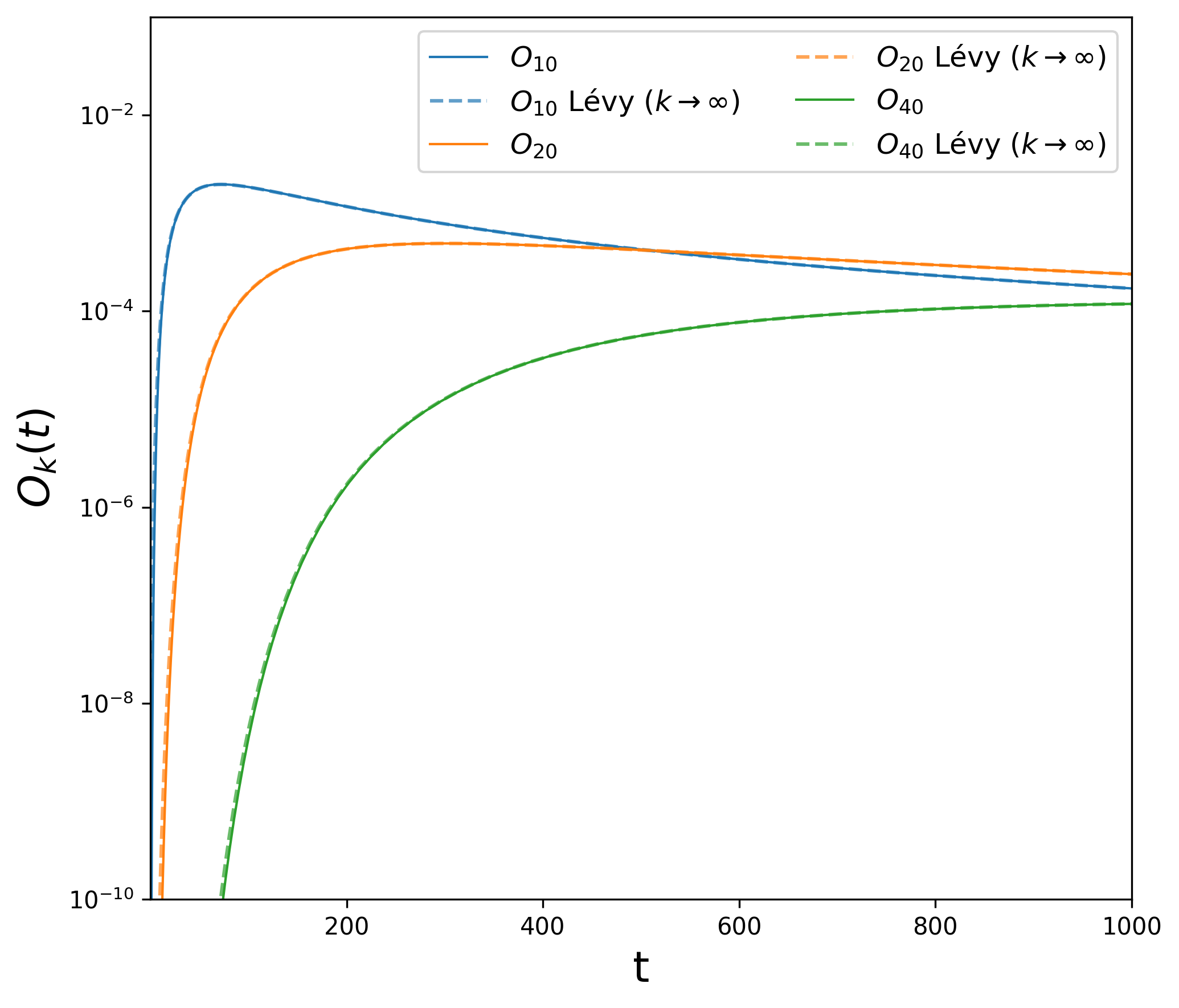}
 \caption{Exact numerical convolution of the $q$-exponential density compared with the shifted stable-Lévy approximation in Eq.~\eqref{eq:stable-levy-asymptotic} for the critical case $q=5/3$. The upper panel shows early cascade stages ($k=3,4,5$), while the lower panel shows larger cascades ($k=10,20,40$). The agreement improves with increasing $k$, demonstrating convergence of the repeated-reservoir response toward the asymptotic transport kernel.}
\label{fig:conv-vs-analytical}
\end{figure*}

For large $k$, an analytical approximation can be obtained from the small-$s$ expansion. For $q=5/3$ we have
\begin{equation}
\tilde{\psi}_{5/3}(s) = 1 - \sqrt{\pi z} e^z \operatorname{erfc}(\sqrt{z})\,,
\end{equation}
where $z\equiv\frac{3s}{2\lambda}$. In the low-frequency regime $s\ll1$, expansion up to the required order gives:
\begin{equation}
\tilde{\psi}_{5/3}(z) \approx 1 - \sqrt{\pi} z^{1/2} + 2z\,.
\end{equation}
The expression for $\ln{\tilde{\psi}^k}=k\ln{\tilde{\psi}_{5/3}}$ becomes
\begin{equation}
\ln \tilde{\psi}^k(s) \approx -k \sqrt{\pi \frac{3s}{2\lambda}} + k \left( 2 - \frac{\pi}{2} \right) \frac{3s}{2\lambda}=-k \sqrt{\pi \frac{3s}{2\lambda}}+\frac{3(4-\pi)}{4\lambda}s\,.
\end{equation}
The linear term in $s$ corresponds to a time shift, so we write
\begin{equation}
\tilde{\psi}^k(s) =e^{-k \sqrt{ \frac{3\pi s}{2\lambda}}+\Delta_k \,s}\,.
\label{eq19}
\end{equation}
Using the inverse Laplace transform, the large-$k$ outflow is approximated by the following $\alpha$-stable Lévy density:
\begin{equation}
O_k\sim \frac{k}{2}\sqrt{\frac{3}{2\lambda (t+\Delta_k)^3}} e^{-\frac{3\pi k^2}{8\lambda (t+\Delta_k)}}\,, \label{eq:stable-levy-asymptotic}
\end{equation}
where $\Delta_k\equiv\frac{3k(4-\pi)}{4\lambda}$. FIG~\ref{fig:conv-vs-analytical} illustrates this critical case and shows how the shifted stable-Lévy approximation improves as the number of reservoirs increases, confirming that the asymptotic expression becomes increasingly accurate along the cascade.

\subsection*{The continuous limit for the outflow field}

To describe the transition from the discrete reservoir cascade to a continuous space-time description, we establish a scaling limit. Let $x_k = k \Delta x$ represent the continuous spatial coordinate, where $k$ is the reservoir index and $\Delta x$ is the characteristic spatial step. As the number of cascading elements tends to infinity ($k \to \infty$) and the spatial increment vanishes ($\Delta x \to 0$), the discrete stage parameters are mapped onto macroscopic scales.

For the critical  index $q = 5/3$, the low-frequency asymptotics derived in Eq. \eqref{eq19}  imply that the Laplace transform of the outflow at position $x$ has the stable-Lévy form~\cite{Feller1971,Zolotarev1986}:
\begin{equation}
\tilde{O}(x,s) \approx \exp\left( - \frac{x}{\Delta x} \sqrt{\frac{3\pi s}{2\lambda}} + \frac{3x(4-\pi)}{4\lambda \Delta x} s \right)\,.
\end{equation}
We define the macroscopic physical parameters of the continuous medium as $D_0 \equiv 2\lambda(\Delta x)^2 / (3\pi)$ (anomalous dispersion coefficient) and $v \equiv 4\lambda \Delta x / [3(4-\pi)]$ (effective advective transport velocity), which yields:
\begin{equation}
\tilde{O}(x,s) \approx \exp\left( -\frac{x}{\sqrt{D_0}}\sqrt{s} + \frac{x}{v}s \right)\,.
\end{equation}
Applying the inverse Laplace transform with the stable-Lévy kernel in conjunction with the time-shift theorem induced by the advective linear term, we obtain the continuous outflow hydrograph in the time domain:
\begin{equation}
O(x,t) \approx \frac{x}{2\sqrt{\pi D_0 \left(t + \frac{x}{v}\right)^3}} \exp\left( -\frac{x^2}{4 D_0 \left(t + \frac{x}{v}\right)} \right)
\end{equation}
where the formulation is restricted to the physical domain $t \ge 0$. Since both the macroscopic coordinate $x$ and the effective advective velocity $v$ are strictly positive, the causality condition $t + x/v > 0$ is naturally satisfied for all physical times. Consequently, the Heaviside step function trivially evaluates to unity ($\Theta = 1$) and can be safely omitted from the analytical hydrograph expression, using the step function wherever formal boundaries are required. For long-term asymptotic behavior ($t \to \infty$ and $t \gg x/v$), the exponential factor approaches unity, and the field decays as a power law governed by the super-diffusive tail: $O(x,t) \propto t^{-1.5}$.

\subsection*{Analytical expression for the storage density}

To determine the continuous fluid volume distribution, we apply the mass conservation law in the Laplace domain. Defining the continuous storage density per unit length as $S(x,t) \equiv S_j(t)/\Delta x$, the continuum limit maps the discrete spatial difference onto a continuous spatial derivative:
\begin{equation}
\tilde{S}(x,s) = -\frac{1}{s} \frac{\partial \tilde{O}(x,s)}{\partial x}\,.
\end{equation}
Differentiating the macroscopic transfer function $\tilde{O}(x,s) = \exp\left( -\frac{x}{\sqrt{D_0}}\sqrt{s} + \frac{x}{v}s \right)$ with respect to $x$ yields:
\begin{equation}
\frac{\partial \tilde{O}(x,s)}{\partial x} = \left( -\frac{\sqrt{s}}{\sqrt{D_0}} + \frac{s}{v} \right) \tilde{O}(x,s)\,.
\end{equation}
Substituting this relation back into the continuous storage equation gives:
\begin{equation}
\tilde{S}(x,s) = \left( \frac{1}{\sqrt{D_0 s}} - \frac{1}{v} \right) \tilde{O}(x,s) = \tilde{S}_A(x,s) - \tilde{S}_B(x,s)\,,
\label{eq:exact_Q_laplace}
\end{equation}
where the two spectral components are defined as:
\begin{align}
\tilde{S}_A(x,s) &= \frac{1}{\sqrt{D_0}} \frac{\exp\left( -\frac{x}{\sqrt{D_0}}\sqrt{s} + \frac{x}{v}s \right)}{\sqrt{s}}\,, \\
\tilde{S}_B(x,s) &= \frac{1}{v} \exp\left( -\frac{x}{\sqrt{D_0}}\sqrt{s} + \frac{x}{v}s \right)\,.
\end{align}

To analytically invert Eq.~\eqref{eq:exact_Q_laplace} while strictly preserving causality, we utilize the time-shifted Laplace pairs for anomalous diffusion kernels:
\begin{align}
\mathcal{L}^{-1}\left\{ \frac{e^{-a\sqrt{s}} e^{bs}}{\sqrt{s}} \right\} &= \frac{1}{\sqrt{\pi (t+b)}} \exp\left(-\frac{a^2}{4(t+b)}\right) \Theta(t+b)\,, \\
\mathcal{L}^{-1}\left\{ e^{-a\sqrt{s}} e^{bs} \right\} &= \frac{a}{2\sqrt{\pi (t+b)^3}} \exp\left(-\frac{a^2}{4(t+b)}\right) \Theta(t+b)\,,
\end{align}
where $a = x/\sqrt{D_0} > 0$, $b = x/v > 0$, and $\Theta(\cdot)$ represents the Heaviside step function. Evaluating these inverse transforms for the individual components yields:
\begin{align}
S_A(x,t) &= \frac{1}{\sqrt{\pi D_0 \left(t + \frac{x}{v}\right)}} \exp\left( -\frac{x^2}{4 D_0 \left(t + \frac{x}{v}\right)} \right) \Theta\left(t + \frac{x}{v}\right), \\
S_B(x,t) &= \frac{x}{2 v \sqrt{\pi D_0 \left(t + \frac{x}{v}\right)^3}} \exp\left( -\frac{x^2}{4 D_0 \left(t + \frac{x}{v}\right)} \right) \Theta\left(t + \frac{x}{v}\right).
\end{align}
Subtracting $S_B(x,t)$ from $S_A(x,t)$ provides the analytical solution for the continuous storage density:
\begin{equation}
S(x,t) = \frac{1 - \frac{x}{2 v \left(t + \frac{x}{v}\right)}}{\sqrt{\pi D_0 \left(t + \frac{x}{v}\right)}} \exp\left( -\frac{x^2}{4 D_0 \left(t + \frac{x}{v}\right)} \right) \Theta\left(t + \frac{x}{v}\right).
\label{eq:exact_Q_time}
\end{equation}
Within the physical domain restricted to $t \ge 0$, the condition $t + x/v > 0$ is satisfied for all macroscopic coordinates $x > 0$ and forward velocities $v > 0$. Under these conditions, the Heaviside operator identically equals unity ($\Theta = 1$), meaning the operational dynamics are governed entirely by the continuous algebraic factor.

\subsection*{Storage normalization and conservation}

We evaluate mass conservation for both the outflow field and the total system storage using the boundary conditions of the continuous model.

\subsubsection*{Normalization of the outflow field}

The cumulative volume of fluid passing through a continuous boundary cross-section $x$ over the entire timeline corresponds to the low-frequency limit ($s \to 0$) of the Laplace-transformed hydrograph:
\begin{equation}
S_O = \int_{0}^\infty O(x,t) \,\mathrm{d}t = \lim_{s \to 0} \tilde{O}(x,s)\,.
\end{equation}
Substituting the continuous transfer function $\tilde{O}(x,s) = \exp\left( -\frac{x}{\sqrt{D_0}}\sqrt{s} + \frac{x}{v}s \right)$ into the limit yields:
\begin{equation}
S_O = \lim_{s \to 0} \exp\left( -\frac{x}{\sqrt{D_0}}\sqrt{s} + \frac{x}{v}s \right) = \exp(0) = 1\,,
\end{equation}
which satisfies the temporal flux normalization condition.

\subsubsection*{Total storage conservation}

The total storage inside the open spatial domain $x \in [0, \infty)$ at any given physical time $t > 0$ is defined as:
\begin{equation}
\mathcal{S}(t) = \int_0^\infty S(x,t) \,\mathrm{d}x\,.
\end{equation}
Using the continuous spatial derivative identity, the total mass in the Laplace domain is expressed as:
\begin{equation}
\tilde{\mathcal{S}}(s) = \int_0^\infty \tilde{S}(x,s) \,\mathrm{d}x = -\frac{1}{s} \int_0^\infty \frac{\partial \tilde{O}(x,s)}{\partial x} \,\mathrm{d}x\,.
\end{equation}
By the Fundamental Theorem of Calculus, this spatial integral evaluates to:
\begin{equation}
\tilde{\mathcal{S}}(s) = -\frac{1}{s} \left[ \tilde{O}(\infty, s) - \tilde{O}(0,s) \right]\,.
\end{equation}
Given the transfer function $\tilde{O}(x,s) = \exp\left[ -\left( \frac{\sqrt{s}}{\sqrt{D_0}} - \frac{s}{v} \right) x \right]$, we evaluate the boundary limits:
\begin{enumerate}
    \item At $x = 0$:
    \begin{equation}
    \tilde{O}(0,s) = \exp(0) = 1\,.
    \label{eq:boundary_0}
    \end{equation}
    \item As $x \to \infty$: for any complex frequency $s$ in the right half-plane where $\text{Re}(\sqrt{s}) > 0$, the spatial decay of the anomalous diffusion term dominates the advective linear growth. Thus:
    \begin{equation}
    \tilde{O}(\infty, s) = \lim_{x\to\infty} \exp\left[ -\left( \frac{\sqrt{s}}{\sqrt{D_0}} - \frac{s}{v} \right) x \right] = 0\,.
    \label{eq:boundary_inf}
    \end{equation}
\end{enumerate}
Substituting Eqs.~\eqref{eq:boundary_0} and \eqref{eq:boundary_inf} into the mass relation yields:
\begin{equation}
\tilde{\mathcal{S}}(s) = -\frac{1}{s} [0 - 1] = \frac{1}{s}\,.
\end{equation}
Applying the inverse Laplace transform, we find:
\begin{equation}
\mathcal{S}(t) = \mathcal{L}^{-1}\left\{ \frac{1}{s} \right\} = \Theta(t) = 1\,, \quad \forall t \ge 0\,.
\end{equation}
This confirms that the continuous storage density $S(x,t)$ maintains the conservation of the stored volume over time.

\section*{Final remarks}

The primary analytical outcome of this work is the derivation of the large-cascade asymptotic flow rate for the critical index $q=5/3$. In this regime, the response of the reservoir chain converges strictly to an $\alpha$-stable L\'{e}vy density whose temporal argument incorporates a Galilean-type transformation shift. This linear shift arises from the regular leading-order term in the low-frequency expansion of the Laplace transform, encoding an effective advective displacement superposed on the underlying anomalous, heavy-tailed spreading. The resulting shifted L\'{e}vy hydrograph provides a compact, closed-form macroscopic description of the long-cascade response while rigorously preserving the structural mass balance inherited from the discrete stage-wise equations.

These findings are  relevant to a broad class of natural and engineering transport processes characterized by structural heterogeneity, broad waiting-time spectra, and non-exponential relaxation scales. Potential applications span watershed hydrographs, solute transport in disordered or fractured porous media, contaminant migration in aquifers, residence time distributions in chemical and biochemical reactors, non-ideal mixing in process units, and anomalous heat or particle transport in complex materials. For these systems, the exact convolutional framework offers a physically constrained approach to modeling heavy-tailed scaling without violating conservation principles, while the shifted L\'{e}vy limit provides an analytically tractable macroscopic kernel for high-order cascades.

Our analysis clarifies how nonextensive statistical mechanics connects to operational transport models by distinguishing between two distinct physical paradigms: ensemble parameter fluctuations and sequential renewal dynamics. A prominent example is the relationship between the Erlang distribution and the $q$-generalized gamma distribution, also known in this context as the $F$-distribution. This distribution was introduced empirically~\cite{Queiros2007} and subsequently derived from the fluctuations of intensive parameters within the framework of Beck-Cohen superstatistics~\cite{Beck2001,Beck2003,Sattin2004,Queiros2005}. This formulation has successfully modeled non-Gaussian fluctuations in financial trading volumes~\cite{Queiros2005} and granular-gas dynamics~\cite{Sattin2004}. However, because its physical foundation lies in superstatistical ensemble averages rather than sequential transport, the $q$-generalized gamma distribution is structurally distinct from the $k$-fold convolution of a $q$-exponential kernel. This distinction is crucial for transport theory: convolution is strictly dictated by sequential renewal operations and mass conservation across a chain of elements, whereas the parametric deformation represents an integration over fluctuating local environments.

We have demonstrated that replacing the exponential factor in the classical Erlang/gamma law with a $q$-exponential function does not yield a distribution equivalent to the repeated convolution of $q$-exponential waiting times. This distinction is physically fundamental rather than merely technical. In multi-stage cascade models, convolution represents the operational consequence of mass conservation through sequential elements, tying the transport dynamics directly to renewal structures. While the $q$-generalized gamma distribution provides a valuable description for specialized nonextensive media, it represents a distinct physical architecture compared to explicit convolutional cascades where mass progresses through sequential storage stages.

From a mathematical perspective, our results establish a clear connection between the $k$-fold convolution of $q$-exponentials and $\alpha$-stable L\'evy distributions, thereby bridging two distinct theoretical frameworks for generating power-law decays.

\section*{acknowledgments}
We acknowledge fruitful discussions with M. A. Pires and C. Tsallis, as well as partial financial support from Conselho Nacional de Desenvolvimento Científico e Tecnológico
 (CNPq) and Coordenação de Aperfeiçoamento de Pessoal de Nível Superior (CAPES) (Brazilian agencies). This article was produced with the support of the Erasmus+ project 2024-1-HU01-KA131-HED-000209516.


\begin{thebibliography}{99}
\bibitem{Nash1957}
J. E. Nash, \textit{The form of the instantaneous unit hydrograph}, Proceedings of the General Assembly of Toronto, International Association of Scientific Hydrology \textbf{3}, 114--121 (1957).

\bibitem{Nash1959}
J. E. Nash, \textit{Systematic determination of unit hydrograph parameters}, Journal of Geophysical Research \textbf{64}, 111--115 (1959).

\bibitem{Dooge1959}
J. C. I. Dooge, \textit{A general theory of the unit hydrograph}, Journal of Geophysical Research \textbf{64}, 241--256 (1959).

\bibitem{RodriguezIturbe1979}
I. Rodr{\'i}guez-Iturbe and J. B. Vald{\'e}s, \textit{The geomorphologic structure of hydrologic response}, Water Resources Research \textbf{15}, 1409--1420 (1979).

\bibitem{SzollsiNagy1982}
A. Sz{\"o}ll{\H{o}}si-Nagy, \textit{The discretization of the continuous linear cascade by means of state space analysis}, Journal of Hydrology \textbf{58}, 223--236 (1982).

\bibitem{Szilagyi2006}
J. Szilagyi, \textit{Discrete state-space approximation of the continuous Kalinin--Milyukov--Nash cascade of noninteger storage elements}, Journal of Hydrology \textbf{328}, 132--140 (2006).

\bibitem{Singh1997}
V. P. Singh, \textit{Kinematic Wave Modeling in Water Resources: Environmental Hydrology} (John Wiley \& Sons, 1997).

\bibitem{Singh1976}
V. P. Singh and D. A. Woolhiser, \textit{A nonlinear kinematic wave model for watershed surface runoff}, Journal of Hydrology \textbf{31}, 221--243 (1976).

\bibitem{SinghWoolhiser2002}
V. P. Singh and D. A. Woolhiser, \textit{Mathematical modeling of watershed hydrology}, Journal of Hydrologic Engineering \textbf{7}, 270--292 (2002).

\bibitem{Tsallis1988}
C. Tsallis, \textit{Possible generalization of Boltzmann--Gibbs statistics}, Journal of Statistical Physics \textbf{52}, 479--487 (1988).

\bibitem{Tsallis2023}
C. Tsallis, \textit{Introduction to Nonextensive Statistical Mechanics: Approaching a Complex World} (Springer, New York, 2023).

\bibitem{Borges2004}
E. P. Borges, \textit{A possible deformed algebra and calculus inspired in nonextensive thermostatistics}, Physica A: Statistical Mechanics and its Applications \textbf{340}, 95--101 (2004).

\bibitem{Tsallis1995}
C. Tsallis, S. V. F. Levy, A. M. C. Souza, and R. Maynard, \textit{Statistical-mechanical foundation of the ubiquity of L{\'e}vy distributions in nature}, Physical Review Letters \textbf{75}, 3589--3593 (1995).

\bibitem{Umarov2008}
S. Umarov, C. Tsallis, and S. Steinberg, \textit{On a q-central limit theorem consistent with nonextensive statistical mechanics}, Milan Journal of Mathematics \textbf{76}, 307--328 (2008).

\bibitem{Lutz2003}
E. Lutz, \textit{Anomalous diffusion and Tsallis statistics in an optical lattice}, Physical Review A \textbf{67}, 051402 (2003).

\bibitem{LimaCurado2026}
H. S. Lima and E. M. F. Curado, \textit{Emergent heavy-tailed distributions from a Markovian random walk}, arXiv preprint arXiv:2605.22933 (2026).

\bibitem{Soares2025}
S. M. Soares, L. Squillante, H. S. Lima, C. Tsallis, and M. de Souza, \textit{Universally nondiverging Gr\"uneisen parameter at critical points}, Physical Review B \textbf{111}, L060409 (2025).

\bibitem{LimaTsallis2023}
H. S. Lima and C. Tsallis, \textit{Ising chain: Thermal conductivity and first-principle validation of Fourier's law}, Physica A: Statistical Mechanics and its Applications \textbf{628}, 129161 (2023).

\bibitem{LimaTsallis2026}
H. S. Lima, C. Tsallis, D. Eroglu, and U. Tirnakli, \textit{Fourier's law breakdown for the planar-rotor chain with long-range interactions}, Physica A: Statistical Mechanics and its Applications \textbf{685}, 131314 (2026).



\bibitem{LimaPereira2026}
H. S. Lima, R. M. Pereira, L. Moriconi, K. R. Sreenivasan, and C. Tsallis, \textit{Superstatistical Approach to Turbulent Circulation Fluctuations}, arXiv preprint arXiv:2604.15277 (2026).

\bibitem{Nazari2021}
B. Nazari, D.-J. Seo, and S. Kim, \textit{Analytical solution for nonlinear hydrologic routing with general power-law storage function}, Journal of Hydrologic Engineering \textbf{26}, 04021016 (2021).

\bibitem{SandorHontiLima2026}
B. S{\'a}ndor, M. Honti, and H. S. Lima, \textit{A spectral model of power-law decay in natural and engineered systems}, arXiv preprint arXiv:2606.08342 (2026).

\bibitem{Plastino1995}
A. R. Plastino and A. Plastino, \textit{Non-extensive statistical mechanics and generalized Fokker-Planck equation}, Physica A \textbf{222}, 347--354 (1995).

\bibitem{Borland1998}
L. Borland, \textit{Microscopic dynamics of the nonlinear Fokker-Planck equation: A phenomenological model}, Physical Review E \textbf{57}, 6634--6642 (1998).

\bibitem{Bologna2000}
M. Bologna, C. Tsallis, and P. Grigolini, \textit{Anomalous diffusion associated with nonlinear fractional derivative Fokker-Planck-like equation: Exact time-dependent solutions}, Physical Review E \textbf{62}, 2213--2218 (2000).

\bibitem{Olver2010}
F. W. J. Olver, D. W. Lozier, R. F. Boisvert, and C. W. Clark, \textit{NIST Handbook of Mathematical Functions} (Cambridge University Press, 2010).

\bibitem{Landau1944} 
L. Landau, \textit{On the energy loss of fast particles by ionization}, J. Phys. USSR \textbf{8}, 201--205 (1944).

\bibitem{Chambers1976}
J.~M.~Chambers, C.~L.~Mallows, and B.~W.~Stuck, \textit{A method for simulating stable random variables}, J. Amer. Statist. Assoc., vol.~\textbf{71}, ~340--344 (1976).

\bibitem{Nolan1997}
J.~P.~Nolan, \textit{Numerical calculation of stable densities and distribution functions},Comm. Statist. Stochastic Models, vol.~\textbf{13}, ~759--774 (1997).

\bibitem{Gorska2011}
K.~Gorska and K.~A.~Penson, \textit{L\'evy stable two-sided distributions: exact and explicit densities for asymmetric case}, Phys. Rev. E, vol.~\textbf{83}, ~061125 (2011).

\bibitem{Feller1971}
W. Feller, \textit{An Introduction to Probability Theory and Its Applications, Volume II} (John Wiley \& Sons, 1971).

\bibitem{Zolotarev1986}
V. M. Zolotarev, \textit{One-Dimensional Stable Distributions} (American Mathematical Society, 1986).

\bibitem{Queiros2007}
S. M. Duarte Queir{\'o}s, L. G. Moyano, J. de Souza, and C. Tsallis, \textit{A nonextensive approach to the dynamics of financial observables}, The European Physical Journal B \textbf{55}, 161--167 (2007).

\bibitem{Beck2001}
C. Beck, \textit{Dynamical foundations of nonextensive statistical mechanics}, Physical Review Letters \textbf{87}, 180601 (2001).

\bibitem{Beck2003}
C. Beck and E. G. D. Cohen, \textit{Superstatistics}, Physica A: Statistical Mechanics and its Applications \textbf{322}, 267--275 (2003).

\bibitem{Sattin2004}
F. Sattin, \textit{Superstatistics from a different perspective}, Physica A: Statistical Mechanics and its Applications \textbf{338}, 437--444 (2004).

\bibitem{Queiros2005}
S. M. Duarte Queir{\'o}s, \textit{On non-Gaussianity and dependence in financial time series: a nonextensive approach}, Quantitative Finance \textbf{5}, 475--487 (2005).

\end{thebibliography}
\end{document}